\documentclass[11pt]{article}
\usepackage{moriond,epsfig}

\def\cf{{cf.~}}

\def\lsim{\raise0.3ex\hbox{$<$}\kern-0.75em{\lower0.65ex\hbox{$\sim$}}} 
\def\gsim{\raise0.3ex\hbox{$>$}\kern-0.75em{\lower0.65ex\hbox{$\sim$}}} 
\def\sc1{\raise2.1ex\hbox{\tiny $r\!\!=\!\!4$}\kern-0.95em{\hbox{$=$}}}

\def\cm3{~{\rm cm^{-3}}}

\def\hinv{$h^{-1}$}

\def\ltsima{$\; \buildrel < \over \sim \;$}
\def\simlt{\lower.5ex\hbox{\ltsima}}
\def\gtsima{$\; \buildrel > \over \sim \;$}
\def\simgt{\lower.5ex\hbox{\gtsima}}

\def\sc{{\rm Science\ }}

\newcommand{\gr}{$\gamma$-ray }
\newcommand{\epm}{e$^\pm$ }
\newcommand{\cre}{e$^-$ }
\newcommand{\pnd}{$\pi^0$-decay }

\newcommand{\ic}{IC }

\def\be{\begin{equation}}
\def\ee{\end{equation}}
\def\bea{\begin{eqnarray}}
\def\eea{\end{eqnarray}}

\begin{document}
\vspace*{4cm}
\title{Gamma-rays from the Large Scale Structure of the Universe}

\author{Francesco Miniati}

\address{Max-Planck-Institut f\"ur Astrophysik,
     Karl-Schwarzschild-Str. 1, 85740, Garching, Germany}

\maketitle\abstracts{
Gamma-ray astronomy will play a crucial role in
the investigation of nonthermal 
processes in the large scale structure of the universe.
Particularly, galaxy clusters (GC) observations at this photon 
energy will help us understand the
the origin of radio emitting high energy particles,
the possible level of cosmic-ray (CR) pressure
in intracluster environment, 
and the strength of intracluster magnetic fields.
In addition here we point out the importance of these 
observations for a possible detection of cosmic shocks
through \gr emission and for 
constraining their CR acceleration efficiency.
We model spatial and spectral properties of 
nonthermal \gr emission due to shock accelerated 
CRs in GC and {\it emphasize} the
importance of imaging capability of upcoming
\gr facilities for a correct interpretation of any 
observational results.
}

\section{Introduction}

Cosmic shocks emerge during structure formation in the
universe, as gravitationally driven gas infall onto 
collapsing objects becomes supersonic. 
The dissipation of kinetic energy at these shocks
raises the temperature of the intergalactic gas. In 
galaxy clusters (GC), characterized by accretion velocities 
of order $\sim 10^3$ km s$^{-1}$, the gas temperature reaches
$\sim 10^8$K. Also, according to recent numerical simulations
most of the low redshift baryons, observed at high redshift as 
Ly-$\alpha$ absorbers have been shock heated to a temperature 
of $10^5-10^7$ K \cite{ceos99a}.

Astrophysical shocks are collisionless and it is believed that
as part of the dissipation mechanism supra-thermal populations
of particles (cosmic-rays, CR hereafter) 
are generated via first order Fermi mechanism \cite{blei87}.
Both CR protons and electrons can be accelerated. 
CR electrons are characterized by rather short radiative lifetime.
For a typical intra-cluster environment, the main 
loss mechanism is inverse Compton (IC) for energies above 
$\sim 150$ MeV and Coulomb losses below that.
Low energy, sub-relativistic CR protons also suffer Coulomb losses
as electrons of the same energy. However, for relativistic
protons the dominant energy loss mechanism 
up to the energy threshold for photo-pair production
is p-p inelastic collisions with a timescale longer than 
a Hubble time. Therefore, once accelerated CR protons up to 
$\sim 10^{15}$ eV accumulate within large scale structures where
they can be confined by $\mu$G strong turbulent 
magnetic fields \cite{voahbr96}.

It is of interest to investigate the acceleration of CRs at
cosmic shocks for a number of reasons. 
Firstly, GCs exhibit non-thermal radiation.
This mainly consists of diffuse radio emission extending on 
Mpc scales. It is thought to be synchrotron radiation 
but there is no consensus as to the origin of the emitting 
relativistic electrons. 
In addition, excess of radiation with respect 
to thermal emission has been reported at both hard X-ray 
\cite{fufeetal99} and, although more controversially, 
for the extreme ultraviolet part of the spectrum 
\cite{lieuetal96b,bobeko99}.
Secondly, if shocks acceleration operates efficiently,
the proton component could bear a significant fraction of the 
total gas pressure, affecting the dynamics of cosmic structures 
\cite{mrkj01}.
Finally, CR electrons accelerated at inter-galactic shocks
could contribute a significant fraction of the cosmic \gr background
(CGB) \cite{lowa00}.

In this contribution we address the role of \gr observations
for investigating CR acceleration at large scale structure shocks.
The results presented here are based on a numerical simulation
which is briefly described in \S \ref{num.sec}.
In \S \ref{cgb.sec} we address the contribution of cosmological 
CRs to the CGB and, based on EGRET experimental results, we 
constraint the efficiency of shock acceleration of CR electrons.
In \S \ref{coma.sec} we present the non-thermal spectrum at 
\gr energies for a Coma cluster prototype and in 
\S \ref{disc.sec} we discuss the importance of \gr observations
for investigating physical conditions in intra-cluster medium (ICM)
environment.

\section{Numerical set-up} \label{num.sec}

We perform numerical simulations that model 
simultaneously the formation of the large scale structure 
and the acceleration, transport and energy losses of three CR 
species: namely {\it shock accelerated} (primary) protons and electrons
and secondary \epm generated in p-p inelastic collisions of the 
simulated CR protons and the thermal nuclei of the IGM. 
The simulation is fully described elsewhere \cite{min02b} and here
we shall only summarize the most important aspects of it.

As for the large scale structure
we adopt a canonical, flat $\Lambda$CDM model
with total mass density $\Omega_m=0.3$,
vacuum energy density $\Omega_\Lambda= 1- \Omega_m= 0.7$,
normalized Hubble constant
$h\equiv H_0/100$ km s$^{-1}$ Mpc$^{-1}$ = 0.67,
 baryonic mass density, $\Omega_b=0.04$, and spectral index
and cluster normalization for initial density perturbations, 
$n_s=1$ and $\sigma_8=0.9$ respectively. 

The CR dynamics, including injection at shocks, acceleration,
energy losses/gains and spatial transport, is computed 
numerically through the code COSMOCR \cite{min01,min02b}.
CR protons are injected at shocks according
to thermal leakage prescription. With the adopted parameters, 
a fraction about $10^{-4}$ of the particles crossing the shock 
is injected as CRs.
For shocks with Mach number between 4 and 10, which 
are responsible for most of the heating of the IGM \cite{min02b}, 
this roughly corresponds to converting 30\% of the shock 
ram pressure into CR proton pressure.
The injection of primary CR electrons in the acceleration mechanism 
is here simply modeled by assuming that their number ratio to protons
at relativistic energies be given by a parameter $R_{e/p}$.
Observations \cite{mulletal95,alpego01} typically indicate 
$R_{e/p}\sim 10^{-2}$.
In accord to the test particle limit of diffusive shock acceleration
theory \cite{blei87}, the injected particles are redistributed in 
momentum as a power-law with slope related to the shock Mach 
number, $M$, as: $q=4/(1-M^{-2})$.
Each CR population is passively advected with the flow.
Finally, energy losses/gains of the 
accelerated CRs are followed by solving numerically a ``kinetic'' 
equation written in comoving coordinates and integrated over 
a discrete set of logarithmically equidistant momentum intervals
that define the numerical grid in momentum space \cite{min01}.
We follow CR protons in the momentum range between $0.1$ GeV/c and
$10^6$ GeV/c, and CR electrons and \epm between 15 MeV/c and 20 TeV/c. 

\section{Contribution to the Cosmic Gamma-ray Background}
\label{cgb.sec}
\begin{figure}
\begin{center}
\rule{0cm}{0mm}
\psfig{figure=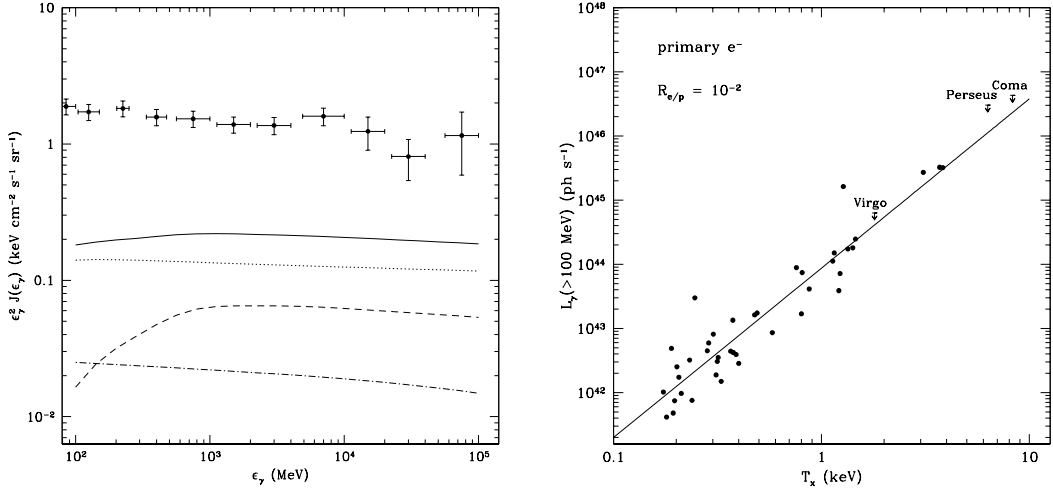,height=7cm}
\caption{{\it Left:} 
Total $\gamma$-ray background flux produced by 
cosmological CRs (solid line), contributions from
IC emission from primary CR electrons (dot line),
$\pi^0$-decay (dash line) and IC from
secondary $e^\pm$ (dash-dot line), and EGRET
experimental data with their error-bars 
(solid dots)$^{13}$ 
{\it Right:} The IC $\gamma$-ray photon luminosity 
above 100 MeV from individual clusters as a function 
of the cluster X-ray core temperature, $T_x$. The flux
is produced by CR electrons accelerated at accretion 
shocks onto GC. EGRET experimental upper limits are 
from Reimer$^{14}$.    
 \label{gamma.fig} }
\end{center}
\end{figure}
In Fig. \ref{gamma.fig} \cite{min02b} (left) we illustrate the 
contribution to the CGB from CRs accelerated at cosmic shocks, 
together with observational data taken from the EGRET experiment
\cite{sreeku98}(solid dots).
The plot shows as a function of photon energy, $\varepsilon$, 
the quantity 
\begin{equation}
\varepsilon^2 J(\varepsilon)
= \varepsilon \, \frac{c}{4\pi H_0} \; \int_{0}^{z_{max}}
\frac{e^{-\tau_{\gamma\gamma}}}{[\Omega_m (1+z)^3 + \Omega_\Lambda]^{1/2}} \;
\frac{j[\varepsilon (1+z),z]}{(1+z)^4} \; dz
\end{equation}
where $j(\varepsilon ,z)$ is the computational-box-averaged
spectral emissivity in units `photons cm$^{-3}$ s$^{-1}$'
computed at each simulation
red-shift, $z$, and at the appropriately blue-shifted photon 
energy $\varepsilon (1+z)$. In addition, $\tau_{\gamma\gamma}$
is an attenuation factor due to photo-pair creation,
$\gamma\gamma \rightarrow e^\pm$, $c$ indicates the speed of 
light and $z_{max}$ an upper limit of integration. 
We consider the following emission processes: IC emission 
of CR electrons scattering off cosmic microwave background 
photons (dot line), decay of neutral pions 
($\pi^0 \rightarrow \gamma\gamma$)
produced in p-p inelastic collisions (dash line) and
\ic emission from secondary \epm (dot-dash line). 
In fig. \ref{gamma.fig} the total flux (solid line) 
corresponds roughly to a constant value at the level of 
0.2 keV cm$^{-2}$ s$^{-1}$ sr$^{-1}$ throughout the spectrum. 
It is dominated by \ic emission from primary 
electrons. A fraction of 
order 30\% and 10 \% is produced by $\pi^0$-decay 
and \ic emission from secondary \epm respectively.

Remarkably, all three components produce the 
same flat spectrum, similar in shape to the observed one.
This result is a reflection of the fact that the CRs
distributions producing the \gr radiation were generated in 
strong shocks with a log-slope $q\simeq 4$ \cite{min02b}.
However, the computed flux is only $\sim$ 15 \% of the 
observed CGB. It is difficult to imagine a higher 
contribution from \pnd (dash line) and IC emission from 
\epm (dot dash line). In fact, if more CR protons were produced
at shocks, CR-induced shock modifications would actually end up
reducing the population of \gr emitting protons (and \epm).
It is possible, however, at least in principle that CR electrons 
would be accelerated more efficiently than assumed here.
In fact, by setting $R_{e/p} = 10^{-2}$ we assume that for 
strong shocks a fraction $\eta \sim 4 \times 10^{-3}
[\log(p_{max}/m_ec)/\log(4\times 10^7)]$ 
($p_{max}$ is the maximum momentum of the accelerated electrons)
of the shock ram pressure is converted into CR electrons.
This is an important matter because if one assumes that $\eta$
is of order of a few percent then the origin of the unexplained
component of the CGB could be accounted for \cite{lowa00}.

However, the electron acceleration efficiency can be constrained by
comparing  the simulated clusters' \gr photon luminosity
above 100 MeV to the upper limits set by the EGRET 
experiment \cite{sreeku96,reimer99} for nearby GCs.
This is done in fig. \ref{gamma.fig} \cite{min02b} (right panel).
The best-fit to the plotted data (solid curve) was found 
through a least-$\chi ^2$ analysis to be:
\begin{equation} \label{gamfit.eq}
{\rm L}_\gamma (>100~ {\rm MeV})= 8.7\times 10^{43}  
\left(\frac{\eta}{4\times 10^{-3}}\right)
\; \left(\frac{\rm T_x}{\rm keV}\right)^{2.6} \; {\rm ph ~s}^{-1}.
\end{equation}
According to our calculations, 
the EGRET upper limits (particularly those
relative to Coma and Virgo clusters) require that 
$\eta \leq 0.8 \%$. For the adopted proton injection efficiency 
this roughly translates into $R_{e/p} \leq 2 \times 10^{-2}$.
In any case, this implies an upper limit on the computed 
\gr flux of about 0.35 keV cm$^{-2}$ s$^{-1}$ sr$^{-1}$.
Thus, according to our computation, cosmological CRs could contribute 
an important fraction of order $\sim 25$ \%  of the CGB.  

\section{\gr for a Coma-cluster prototype: synthetic emission
maps and spectrum}
\label{coma.sec}

In this section we explore the spectral and spatial properties
of non-thermal \gr radiation between 10 keV and 10 TeV due to 
shock accelerated CRs in GC. Clusters
observations in this photon energy range are important for 
a number of issues which we address in detail in the discussion 
section. Among these are the following: correct measurements of 
intra-cluster magnetic field, estimate of CR proton pressure 
component, possible detection of cosmic shocks, estimate 
of CR acceleration efficiency.
However, for a correct interpretation of the observational 
results it is important that the emission produced by different 
particle components be properly identified.
In this respect, the results presented in the following are 
aimed at helping a correct diagnostic of future \gr observations.

As non-thermal bremsstrahlung turns out
unimportant \cite{min02a}, we consider the following emission 
processes: \pnd and IC emission from both primary CR electrons 
and secondary \epm. We first compute the emission spectrum 
for the largest virialized object in the computational box
which has an X-ray core temperature $T_x\simeq 4$ keV. 
The various emission components are then renormalized to the
case of a Coma-like cluster. In particular, IC emission 
from electrons is rescaled according to equation \ref{gamfit.eq}.
The total number of \epm is rescaled assuming that these particles
are responsible the for synchrotron emission of Coma radio halo. For 
the purpose we took a radio flux at 1.4 GHz $S_{1.4GHz} = 640$ mJy 
\cite{deissetal97} and a volume average magnetic field 
$\langle B \rangle\sim 3 \mu$G. \cite{kkdl90}
For the ensuing discussion it is not necessary to assume that
the CRs responsible for the synchrotron emission are
secondary \epm.
We know that these high energy particles exist from the observed radio 
flux and the morphology of radio halos suggest that they reside in 
the inner region of clusters. However, 
assuming that these are secondary \epm allows us to also normalize
the rate of p-p inelastic collisions that determine the total \pnd flux.

The simulation results indicate that
the radiation emitted by primary CR electrons and both \pnd as 
well as secondary \epm originates in spatially separate regions. 
This is a reflection of both the spatial distribution of the 
emitting particles as well as the nature of the emission process. 
Thus, on the one hand, both \epm and $\pi^0$ are produced at the 
highest rate in the densest regions where both the parent CR ions 
and target ICM nuclei are most numerous. Consequently, \ic from \epm
and \pnd emissivities are strongest in the cluster inner regions 
and quickly fade toward its outskirts.
On the other hand,  because of severe energy losses, 
\gr emitting high energy primary electrons are only found 
in the vicinity of strong shocks where they are being accelerated. 
Thus, because the strongest shocks are located at the cluster
outskirts rather than at its core where the ICM temperature is 
high \cite{minetal00}, the spatial distribution of the 
emissivity is now reversed with respect to the previous case. 
\begin{figure}
\rule{-5mm}{0.2mm}
\psfig{figure=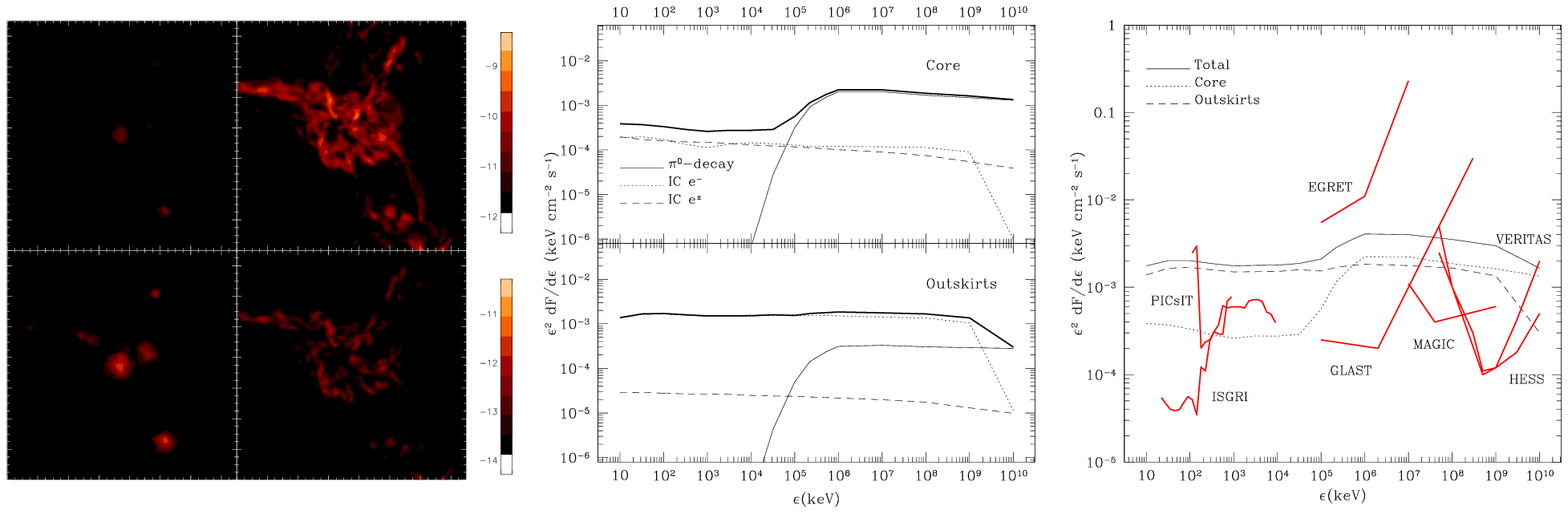,height=6cm}
\caption{{\it Left:} 
Synthetic map of the integrated photon flux 
above 100 keV (top) and 100 MeV (bottom) in units 
``ph cm$^{-2}$ s$^{-1}$ arcmin$^{-2}$'' from
\ic emission by secondary \epm (top-left), 
primary \cre (top-right, bottom-right), and 
\pnd (bottom-left). Each panel measures 15 
\hinv Mpc on a side.
{\it Center:} 
Synthetic spectra from 10 keV up to 10 TeV extracted from a
core (top) and an outskirts region (bottom).
{\it Right:} 
Comparison of synthetic spectra with nominal sensitivity limits 
of future \gr observatories (thick-solid lines). 
For INTEGRAL-IBIS imagers (ISGRI and PICsIT) the curves 
correspond to a detection significance of 3$\sigma$ 
with an observing time of $10^6$ s. All other sensitivity 
plots refer to a 5$\sigma$ significance. EGRET and GLAST 
sensitivities are shown for one year 
of all sky survey and Cherenkov telescopes
(MAGIC, VERITAS and HESS) for 50 hour exposure
on a single source. 
\label{figc4.fig}}
\end{figure}
In the four left panels of fig. \ref{figc4.fig}
we present synthetic maps of the 
integrated photon flux above 100 keV (top panels) and 100 MeV 
(bottom panels). Among these four panels, the
right panels are associated with \ic emission 
from primary electrons and the left panels to 
\ic from secondary \epm (top-left) and \pnd (bottom-left).
As anticipated
the emission from \pnd (bottom left) and \epm (top left) 
is confined to the cluster core. There it creates a diffuse 
halo which rapidly fades with distance from the center.
On the other hand, \ic emission from primary
\cre is distributed over a much more extended area. Moreover, 
it is characterized by a strikingly rich and irregular 
structure. This is a direct reflection of the complex 
``web'' of shocks that reside at the outskirts of galaxy 
clusters \cite{minetal00}.

After generating multifrequency synthetic maps that include 
all the aforementioned emission components, in the central 
panel of fig. \ref{figc4.fig} we have produced synthetic 
spectra extracted from a core (top) and an outskirts 
region (bottom). The core region corresponds to 
an angular size of $1^o$ and the outskirts region is defined as a
ring with inner and outer radii of $0.5^o$ and $1.5^o$ respectively
(0.5$^o \sim $ 1 Mpc at the red-shift of Coma cluster). 
Based on the above findings, the idea is that by looking at 
the cluster core one should see the \pnd and \epm contributions 
whereas by observing the outskirts one should see \ic 
emission from CR electrons accelerated at accretion shocks.
Fig. \ref{figc4.fig} (center) shows that at high photon energy 
($>$ 100 MeV) the spectrum of radiation is indeed dominated 
by $\pi^0$-decay (solid thin line) in the core region (top panel) 
and by \ic emission (dotted line) from primary electrons in the 
outer region (bottom).
Below $\sim$ 100 MeV the flux from the 
outskirts region is still strongly dominated by \ic 
emission from primary \cre. This contribution
is much reduced in the narrower field of view of the core
region. However it is still significant and at the level of
\ic emission from secondary \epm. Because the radiation flux from
these two components depends on the assumed normalizations 
and the actual shock structure subtended 
by the field of view, their almost equality in
fig. \ref{figc4.fig} (center, top panel) is obviously only indicative.
Nevertheless they are expected of comparable intensity. 
The contribution of the primary CR electrons could be estimated 
by measuring its radial dependence in the outer regions, where it
dominates, and then by extrapolating its value for the core region.

\section{Discussion} \label{disc.sec}

In the right panel of fig. \ref{figc4.fig} we plot sensitivity 
limits of planned \gr observatories together with
the total radiation spectra from the core (dotted line) 
and outskirts (dashed line) regions. 
The sensitivity limits are plotted for
point sources and will be somewhat worse for the case of 
clusters given their large angular size. In any case, 
according to this plot, several future experiments should be 
sensitive enough to detect the computed non-thermal emission 
at most photon energies.
In particular the IBIS imager onboard INTEGRAL should readily 
measure the flux between 100 keV and several MeV. 
In addition GLAST and Ch\v{e}renkov telescopes
should be able to detect both the core \gr emission 
from \pnd as well as the \ic flux directly
produced at accretion shocks above 100 MeV and 10 GeV respectively.

These measurement will provide important information about GCs.
First, the extended emission is produced at the location of 
of accretion shocks. Merger shocks have occasionally 
been observed in the core of clusters as relatively weak 
temperature jumps. However, strong accretion shocks have yet 
to be observed {\it directly}. Thus, detection and imaging
of \ic emission from primary electrons would provide an opportunity 
to directly observe GC accretion shocks.

Furthermore, measuring the flux about 100 keV together 
with radio measurements, will allow an estimate of 
(or upper limits on) the ICM magnetic field. 
It is important, however, to separate out the contribution
from the CR electrons accelerated at the outlining shocks.
Because the magnetic field strength is expected to drop at the
cluster outskirts, these electrons likely generate only a weak radio 
emission. Nonetheless, they produce a strong \ic flux which can easily
dominate the total soft \gr emission (\cf fig. \ref{figc4.fig}).
In this case separating out their contribution would be an important 
step for correctly measuring ICM magnetic fields.

One of the most awaited experiments is related to the measurement 
of the \gr flux at and above 100 MeV. This is important in order to 
convalidate or rule out secondary \epm models for radio 
halos in GC
\cite{doen00,blco99,mjkr01}
and in order to estimate the non-thermal CR pressure 
there \cite{mrkj01}. In this perspective several authors 
estimated for nearby clusters 
the \gr flux from $\pi^0$-decay. However, radiation flux from
\ic emission can be comparable \cite{min02a}. Therefore, 
for a correct interpretation of the measurements the 
contributions from these two
components will need to be separated as outlined here.
Estimating the \gr flux from \ic emission due to primary
CR electrons will also be instrumental for addressing 
the contribution of this emission mechanism to the CGB 
\cite{lowa00,min02b,keshetetal02}.

Measurements at photon energies, from
$\sim 10-100$ GeV to $1-10$ TeV, will also be of great interest.
Because the radiated energy 
spectrum is directly connected to the distribution function 
of the emitting particles, measuring the flux at different 
energies will provide information about the acceleration
mechanism. In particular, the observed spectrum could 
differ from our predictions if the accretion shocks were 
modified by CR pressure. Finally the maximum energy of the 
accelerated CR electrons could be inferred from a spectral cut-off,
although only as long as the latter turns out below the energy 
threshold for $\gamma\gamma$ absorption 
($\sim$ TeV).
Given the different environmental 
conditions with respect to supernova remnants, observations of
galaxy clusters offer the interesting possibility for a 
broad exploration of shock acceleration physics.

\section*{Bibliography}

\bibliographystyle{unsrt}
\bibliography{papers,books,proceed}

\end{document}